# A Novel Technique of Noninvasive Hemoglobin Level Measurement Using HSV Value of Fingertip Image

Md Kamrul Hasan[1], Nazmus Sakib[1], Joshua Field[2], Richard R. Love[1] and Sheikh I. Ahamed[1]

*Abstract—* Over the last decade, smartphones have changed radically to support us with mHealth technology, cloud computing, and machine learning algorithm. Having its multifaceted facilities, we present a novel smartphone-based noninvasive hemoglobin (Hb) level prediction model by analyzing hue, saturation and value (HSV) of a fingertip video. Here, we collect 60 videos of 60 subjects from two different locations: Blood Center of Wisconsin, USA and AmaderGram, Bangladesh. We extract red, green, and blue (RGB) pixel intensities of selected images of those videos captured by the smartphone camera with flash on. Then we convert RGB values of selected video frames of a fingertip video into HSV color space and we generate histogram values of these HSV pixel intensities. We average these histogram values of a fingertip video and consider as an observation against the gold standard Hb concentration. We generate two input feature matrices based on observation of two different data sets. Partial Least Squares (PLS) algorithm is applied on the input feature matrix. We observe $R^2$=0.95 in both data sets through our research. We analyze our data using Python OpenCV, Matlab, and R statistics tool.

## I. INTRODUCTION

Hemoglobin (Hb) is an important protein molecule in blood that transfers oxygen to our body tissue. Hemoglobin quantities keep the proper shape of red blood cells [1]. Based on the shape and amount of red blood cells, people suffers from different types of severe diseases such as anemia and sickle cell. For the diagnosis of these diseases, hemoglobin concentration is measured using the invasive way [2] in clinical setup. Despite satisfactory accuracy, patients as well as health care providers experience several issues with invasive approaches. Since invasive techniques need blood samples, this approach is not always recommended for frequent Hb level measurement of premature infants, aging population, pregnant women, sickle cell, and anemia patients. To reduce the frequent clinical test and overcome the patient management, caregivers sometimes do the blood transfusion without proper blood screening. This practice jeopardies the patients' situation that results in high risk as well as increased medical costs in the future [3].

Patients can measure Hb level using noninvasive point of care (POC) tools and send the reports to the medical provider remotely. Regular hemoglobin reports using these POC tools solve the problem regarding travel cost, time, and emergency room (ER) management problems. Noninvasive techniques have been attempted recently though the performance is not up to the mark regarding accuracy, user friendliness, computational complexity etc. Noninvasive hemoglobin level measurement systems have been proposed using different ways. For example, the ratio of hemoglobin to water in blood plasma [4], [5], [6] using a finger probe consisting of multiple wavelengths of LED light, OxyTrue Hb [7], Masimo [2] have been offered as a noninvasive solution. These noninvasive efforts have manifold problems. For example, various wavelengths of LED light to work in visible and infrared spectrum entails higher cost and device complexity. The device interfacing, computational and operational issues make trouble to the end users. As a result, these endeavors works in lab setup and does not provide clinical solution. We found hyperspectral camera and spectrometers based noninvasive Hb measurement solution [8], [9], [10], [11] which manifest high precision, but they are very much expensive to adopt.

The smartphone is a key player in mHealth technology due to is multifaceted aspects like mobility, high-resolutioncamera, smart application, network connectivity to healthcare providers, and so on. Smartphones have been already using as an mHealth POC tool for medicinal information [12], heart-rate monitoring, sleep monitoring, pulmonology, gait detection [13], symptom monitoring system [14], cancer care [15], palliative care [16], eESAS [17] and point-ofcare diagnostics. Having the smartphone-based exploration, E. Wang et al. presented HemaApp [18] that predicts blood hemoglobin concentration. Using different lighting sources, they analyzed the fingertip video data captured using a smartphone. Their lighting setup made the system complex for users to enlighten the fingertip. In addition, the system included external device costs and operational issues to the users.

Emphasizing those challenges, we present a novel smartphone based noninvasive hemoglobin measurement solution. The RGB pixel intensities of the video frame are converted to HSV color space and generate histogram values. We apply a PLS algorithm on the data recorded from 30 sickle cell patients of the BCW, USA. We also use the technique on 30 subjects' fingertip data who have different physiological issues in AmaderGram, Bangladesh. We represent our analyzed result and regression line that suggest that the prediction model can calculate nearly 95% accurate result using this technique. The main contributions of this paper are as follows.

1) Establish a fundamental idea analyzing fingertip image of a severe sickle cell patient who received blood transfusion.

2) Develop a novel smartphone-based method for the

fingertip video collection.

- 3) Generate the input feature matrix using the frequency of HSV pixels of each video frame. Partial Least Squares (PLS) algorithm is used to build the prediction model.

- 4) Analyze the novel approach applying on two different data sets collected from USA and Bangladesh.

II. MOTIVATION

Finger stick and venipuncture blood draw are the most common method for blood sample collection. In the clinical setup, health assistants collect 3mL of blood for a complete blood count (CBC) test. In CBC test, the user's red blood cells (RBC), white blood cells (WBC), and platelets information are found. Optical instruments are commonly used to measure Hb concentration. Besides, invasive POC tools can be used remotely in blood sample tests. For example, Mission Plus [19] and HemoCue [20] have been used for hemoglobin level monitoring. These devices give an accuracy level of 0.89 at a mean accuracy of 0.5 g/dL while compared to the results from the CBC test [21], [22]. These expensive POC tools still operate using users' blood samples.

As part of noninvasive techniques, researchers used optical instruments such as spectrometer, hyperspectral camera for sample data collection to measure hemoglobin level. Both spectrometer and hyperspectral camera are highly expensive though they show appreciable accuracy. Sensitive optical lens and external devices are also very rare in poor and developing countries. Since these devices are very complex in use, it complicates the system to use in remote-health care setup. A couple of researchers used external LED light as supplemental hardware in addition to the smartphone for hemoglobin level measurement [18], [2]. Jeff Thomson used mechanically couples a stethoscope to a 3D printed attachment [23]. In [24], Hongying Zhu and Aydogan Ozcan developed an optics system where it initially illuminates a flow cytometer test strip. Then, the test strip was attached to a smartphone camera to conduct point-of-care blood tests. The optic systems were dedicated to light up the strip with the required wavelength. In this occasion, the smartphone camera functions as a sensor. In these existing research works, the proposed model faces distinct challenges. For example, size, cost, and user-friendliness are not considered in many cases. In addition, finger position from the external device might be misplaced in their research. Test strips have expiration date and need proper training to use the POC device. The optical system of the POC device also may show aberrant behavior after a couple of months. In this situation, smartphone-based noninvasive techniques is an appreciable alternative.

The USA has nearly 270 million mobile phone users where most of them have a smartphone. The number of smartphone users is burgeoning worldwide day by day. The processing, memory and storage capacity of the smartphone gives a new dimension to it. Smartphone companies add new features and sensors to satisfy customer demand in their latest model. Since the size and price of the optical instruments are not feasible for noninvasive setup, we can use a smartphone in remote-health-care-settings. To overcome expense and mobility, in this paper, we propose a smartphone-based noninvasive Hb measurement tool where the image HSV values are used. Based on the capabilities of smartphones, this research output can make a remarkable change in resource-limited countries.

Based on the insights from these recent research works, we aim to reduce the cost of external hardware for hemoglobin level measurement using the fingertip video analysis. The organization of the paper is as follows. Various noninvasive hemoglobin measurement techniques are illustrated in Section III. The pixel extraction, color space conversion, and fundamental ideas are delineated in Section IV. Patient demography, data collection process, input matrix creation, and data analysis methods are described in Section V. The result is presented and discussed in Section VI. The conclusion is given finally in Section VII. We use the term fingertip video or video alternatively in the following sections. The fingertip video is considered as 10-second video in this paper.

III. RELATED WORKS

Over the last decade, noninvasive solutions have been offered for many physiological issues. We have seen noninvasive accomplishments in bilirubin measurement [25]. Here Khalid M. Alabdulwahhab et al proposed an imageprocessing-based tool reckoning the intensity of yellow color. In [26], Christopher G. Scully et al delineated a smartphonebased monitor for several physiological variables varying color signals of a fingertip placed in contact with its optical sensor. In [27], Mathew J. Gregoski et al proposed an android application and analyzed the heart-rates acquired from a Motorola-Droid to ECG and Nonin 9560BT pulse oximeter readings. Then, Walter Karlen et al from the University of British Columbia-Vancouver presented a new method to automatically detect the optimal ROI (Region of Interest) for the image taken by phone camera to extract a pulse waveform [28]. The paper suggested optimal camera settings and showed that the incandescent white balance mode is a good choice for camera-oximetry-applications on the tested mobile phone such as the Samsung Galaxy Ace. In these approaches, researchers used different colors for developing their prediction model toward noninvasive solutions.

The research community is exploring their extensive ideas to focus on noninvasive blood diagnosis. They consider various sites of the human body such as eye conjunctiva, ear lobe, face, lower lips and fingertip for data collection regarding noninvasive physiological measurement. Hyo-Haeng Lee et al. proposed a mobile-based solution for video-based biosignal measurements, such as pulse rate, oxygen saturation, respiration rate, and blood pressure in [29]. They analyzed the video image pixel information to evaluate the information apropos of the user's health status though the result needed improvement. Ali Madooei et al. demonstrated a pilot study to identify and grade skin Erythema in [30]. Here, the author demonstrated the efficiency of their approaches in reproducing clinical assessments as well as outperforming RGB imaging data. In [31],

Jarrel Seah and Jennifer Tang worked on a project named Eyenaemia where a color calibration card was put next to the eyelid of a particular patient. This card was used to evaluate the redness of the underside of their eyelid when flipped over. Their outcome detected the risk of anemia but rarely measured the actual concentration of hemoglobin.

M. Rajendra Kumar et al proposed hemoglobin levels measurement system for resource-limited settings in [32]. In this paper, they captured an image of a drop of blood from a filter strip under certain controlled conditions. After that, they predicted Hb levels implementing an image processing algorithm. They used classification tree and correlation-based approach to classify four levels of Hemoglobin (Class IHb above 12 g/dl, Class II-Hb between 10 and 12, Class III-Hb between 8 and 10, Class IV-Hb below 8) in the data. The blood sample was collected in an invasive way but their confusion matrix was eventually obtained on the testing set with the overall accuracy of 82%, sensitivity of 83%, and specificity of 82%. Vitoantonio Bevilacqua et al introduced a noninvasive approach to predict hemoglobin level cardinally based on the image analysis of a specific conjunctival region [33]. They applied their solution on seventy-seven anemic and healthy patients that showed good correlation between the hemoglobin level from invasive samples and the value predicted by their algorithm. But the conjunctival image collection is not easy for users point of view due to hand and eyelid movement. In 2010-2011, Alam et al showed optical measurement of the blood at the fingertip. They proposed a finger probe with 6 LEDs for noninvasive prediction of Hb concentration [34], [35] that covered multiple wavelengths in red to IR spectrum (630, 660, 680, 770, 880, and 1300 nm) to reckon hemoglobin to water ratio (H/W) in blood plasma. In [36], using a three LED (670, 810, 1300 nm) finger probe, Kraitl et al attained similar results after conducting an extensive study on fortyone subjects. Since these existing methods have cost and operational issues (sometimes not ubiquitously applicable around the world) due to external devices like LED, optical instruments, and extra sensing devices, in this research, our cardinal objective is to mitigate those problems through this novel approach.

IV. BASIC IDEA

In general, pixel intensity of an image frame is presented as height x width x channels where the value of each channel tells us about how many different color pixels are used to show the image. For example, if the channel value is 3 then the image frame is presented as RGB, HSV or any other three channel pixel values. If the channel value is one then the image is a grayscale image. In general, the pixel intensity value for each color ranged from 0 to 255.

*A. RGB and HSV color space*

RGB stands for red, green and blue. An image has this three basic color components and RGB is the most frequently used color space by the researchers. The reason to use RGB color space is due to its simplicity and easyto-implement approaches. But, the RGB color space has a couple of limitations. For example, RGB is nonlinear with visual perception and its color specification is semi-intuitive.

An image can be presented by its hue, saturation, and value (HSV) information. HSV color space is more intuitive than RGB color space. HSV can separate luminance from chromaticity [37]. HSV color space is three dimensional where the vertical axis is named as value, the angle in the range $[0, 2\pi]$ is defined as hue and the depth of the color is called saturation. RGB features fail to determine the color and intensity variation because the RGB clustering can not detect the boundary of an object properly in an image. In [37], the author found higher recall and precision values for HSV color space with respect to RGB features. Since HSV has better color continuity, these features are used in our project for the histogram generation.

*B. Histogram values*

Feature space analysis is used as a common tool for image understanding. Significant features are located in high density regions in color space [38]. Histogram-based approach is seen to use in the feature space identification. The frequency of each color pixel shown in a histogram represents different pixel contributions in a picture. The peak of histogram changes based on the intensity and brightness of an image. For this reason, we consider histogram value of an image as the input feature matrix. Figure 1 shows the histogram of an image where HSV color space information are used. The number of bin is chosen as 256 so that each pixel frequency is observed clearly. The image is selected from a fingertip video of a subject who has hemoglobin level 10.0 gm/dL. Figure 2 presents the closeup view of the histogram where the frequency of pixels is from 0 to 600. Since this low frequency of HSV pixels are not visible in Figure 1, Figure 2 shows the clear picture of this change.

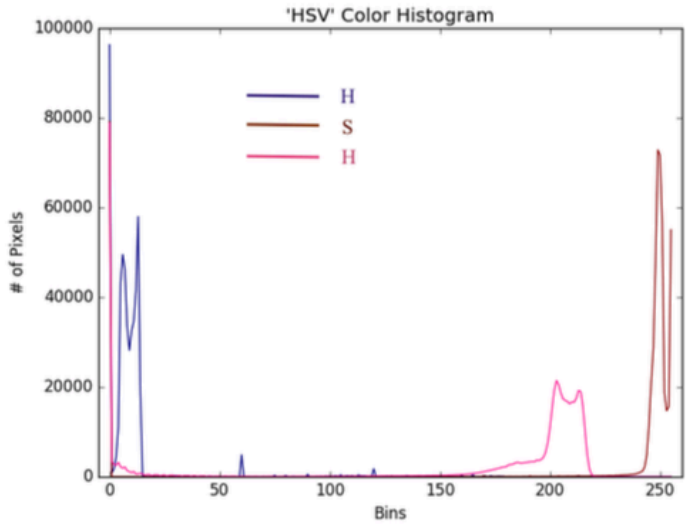

Fig. 1: Frequency of HSV color pixels of an image shown as histogram. The image was captured from a person who had hemoglobin level 10.0 gm/dL.

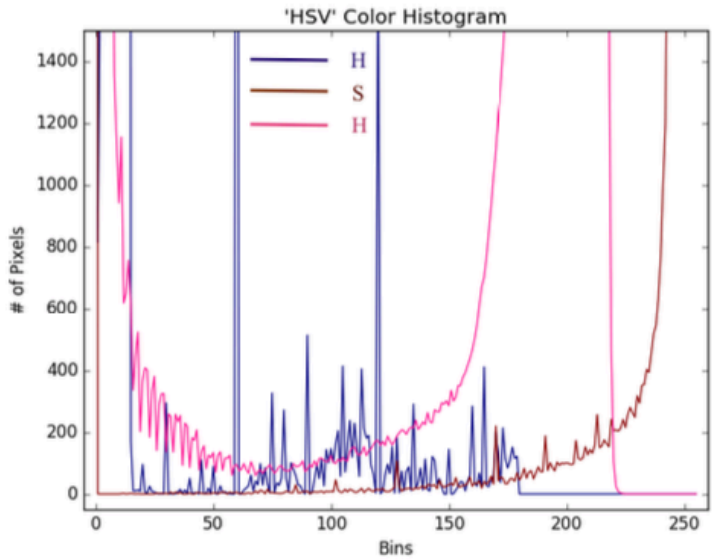

Fig. 2: Closeup view of HSV frequency generated from an image that was captured from a person who had hemoglobin level 10.0 gm/dL.

*C. Fundamental idea*

In this research, we explored the fingertip video image information captured by a smartphone to discover the ground truth about an important bio-marker of human body called hemoglobin. At the beginning of our research, we analyzed the lower and higher level of hemoglobin of a person. We chose one sickle cell patient having severe anemia who received blood transfusion. We recorded a video of that patient before his blood transfusion. The video was captured before his blood sample collection having the patient's consent in the Blood Center of Wisconsin (BCW). This person received a blood transfusion in the BCW on the same day. After two weeks of the blood transfusion, the same person came again to the BCW for regular checkup. The physician advised for his blood diagnosis as part of his physical issues. At that time, we collected the video of that patient before blood sample collection. We got the hemoglobin level report of this subject measured after two weeks of the blood transfusion. The Hb test report showed higher hemoglobin level than that measured before the blood transfusion. We compared these two hemoglobin levels and analyzed the data. We applied image filtering on certain level of HSV pixels of the image. The video captured with lower level Hb (measured before blood transfusion) was analyzed and presented in Figure 4(A). The second fingertip video image, recorded after blood transfusion (higher level of Hb), was also analyzed and presented in Figure 4(B).

We see the fingertip video is reddish when we run the video on a player as shown in Figure 3(A). The above-mentioned sickle cell patient had hemoglobin level 7.5 gm/dL before his blood transfusion. The person was 44 years old when the video was recorded. We converted the RGB pixel of both recorded video images into HSV color space and changed the pixel color using Python OpenCV library as is shown in Figure 4(A). The hemoglobin concentration of this man after blood transfusion was found 10.0 gm/dL. The filtered video image generated from this stage is presented in Figure 4(B). The difference between Figure 4(A) and Figure 4(B) represents distinct change of hemoglobin level of the person. Figure 4(B) has nearly double number of black pixels than that in Figure 4(A). The color difference of these two frames shows a strong intuitive relationship between two different hemoglobin levels and video image. This strong affinity presents a fundamental clue regarding the change of hemoglobin level and image pixel which motivates us to go forward for noninvasive hemoglobin level detection using video image analysis. Based on this ground truth, we come up with a novel HSV-based noninvasive hemoglobin level prediction model in this research paper.

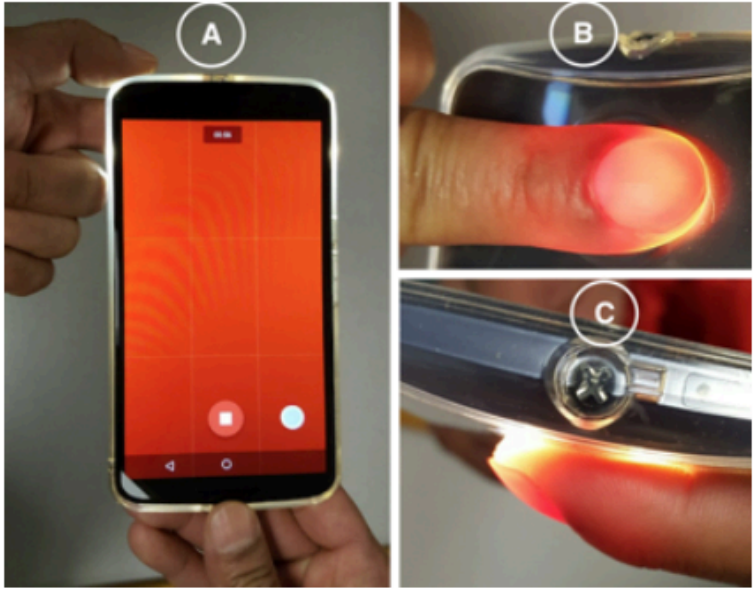

Fig. 3: Fingertip video collection system. A) The reddish video comes from fingertip B) The index finger covers the camera as well as the flash C) The user turns on the flash to enlighten the finger area.

## V. NOVEL APPROACH

Since this novel approach requires only the fingertip video of a patient using the smartphone camera, the approach is out of health hazard and additional hardware cost. The initial step to collect the video sample of a patient is shown in Figure 3. The patient information is given in the following subsection.

### A. Patients demography

We mentioned earlier that we collected the fingertip video in two different places. The first place was the Blood Center of Wisconsin (BCW), USA where we considered 30 subjects as our first pilot study. We found 19 females and 11 male participants in these videos and blood samples collection processes. The data collection process was run for a couple of months since all subjects were not available at the same time. The age range of this patient group was from 19 years to 60 years, and the average was 33. The lowest level of hemoglobin we observed in this group was 6.3 gm/dL, and 12.3 gm/dL was the highest concentration. The demography of this data set is presented in Figure 5. Figure 5 represents that the male subjects have a higher level of Hb than that of the female patients though two men have lower level Hb in this data set. We got only four patients who were older than 50 years as shown in Figure 5. Here round circles are presenting the subjects where we are missing two circles in Figure 5. Since, we get a couple of similar levels of hemoglobin, those two circles are overlapped here.

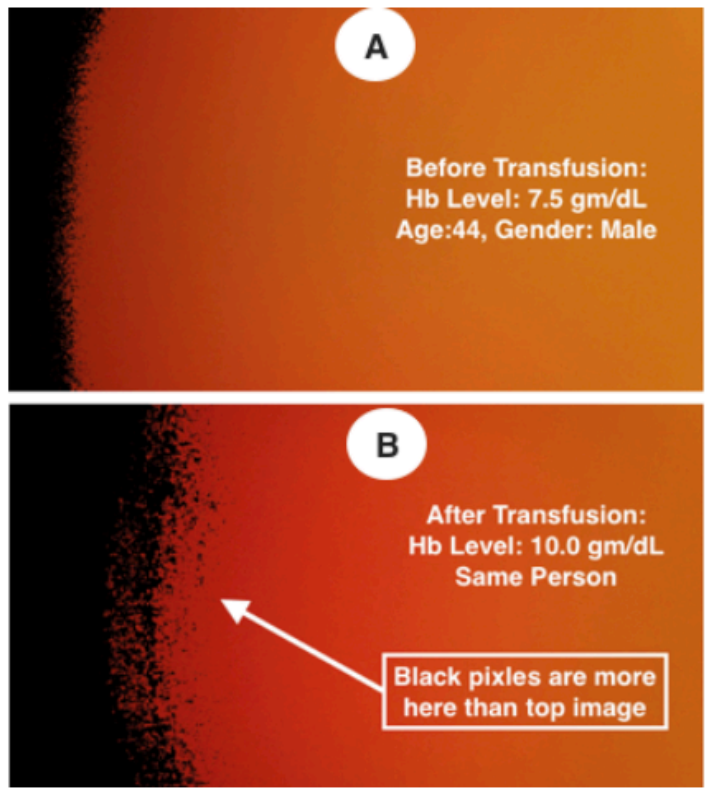

Fig. 4: Video image of a sickle cell patent who received blood transfusion. A) The video was taken before blood transfusion when the Hb level was 7.5 gm/dL B) The fingertip video image of the same person was captured after two weeks of blood transfusion where the Hb level was 10.0 gm/dL.

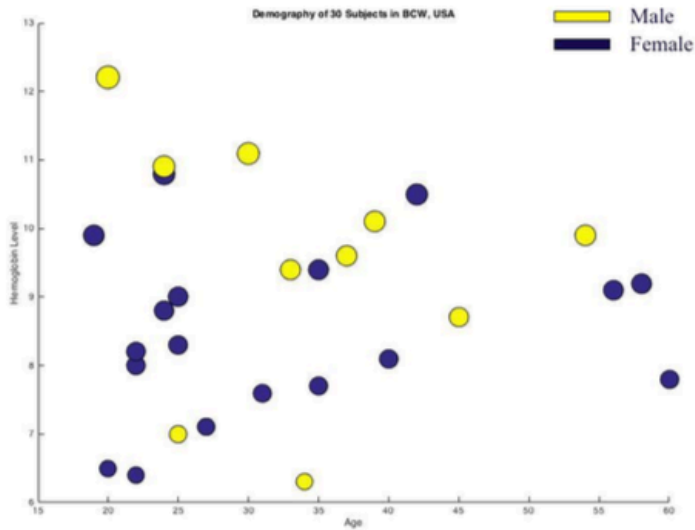

Fig. 5: Demography of the subjects from Blood Center of Wisconsin.

The second location was AmaderGram, Bangladesh that provided 30 patients' fingertip video and respective clinical Hb report. Here twenty females and ten males gave the video and blood sample data. We collected video and blood sample following the similar protocol maintained in the BCW. The data collection process was run for a couple of weeks here.

The age range of this patient group was from 22 to 62 years and the average age was 39. 13 gm/dL was observed as the highest level of hemoglobin, and the lowest value of Hb was found as 7.6 gm/dL here. The demography of this data set is presented in Figure 6. Figure 6 represents that the male subjects have a higher level of Hb than that of the female patients. Here we are missing one female circle in Figure 6 due to circle overlapping for similar levels hemoglobin.

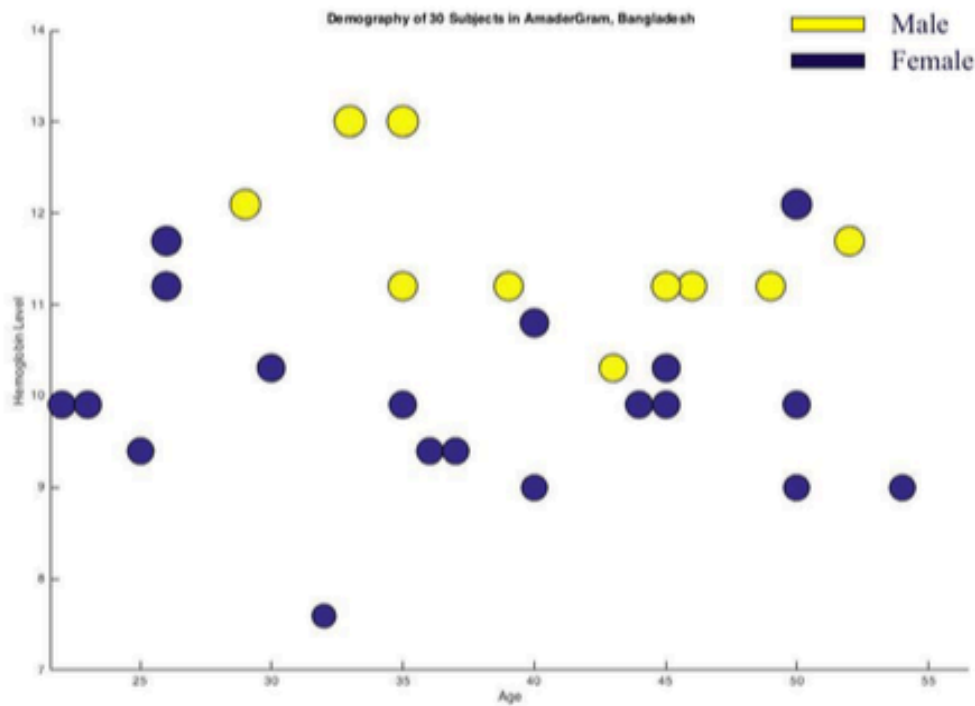

Fig. 6: Demography of AmaderGram, Bangladesh subjects.

*B. Recommendations while data collection*

The fingertip video samples were collected by Google Nexus 4 smartphone. We recommended the following directions while recording the video in both locations.

. Record the fingertip video before blood sample collection

. Use index finger for video recording

. Cover the smartphone camera as well as flash properly so that no ambient light can penetrate as shown in Figure 3(B)

. Turn on the smartphone camera flash while video recording as illustrated in Figure 3(C)

. Record a 10-second video

. Keep record of the name of video file against the clinical Hb report

Collect blood sample with in one hour of video record ing. Test the blood sample in clinical setup and consider as gold standard hemoglobin concentration. *C. Fingertip*

*video processing* In general, a smartphone captures 30 frames per second (fps) of a video. Thus a 10-second video consists of 300 frames that are considered initially in this research. Before color intensity analysis, we checked each video playing one by one. We observed that nearly the first 50 frames contained mostly noise (black/very dark area) for 35% of the total number of videos. We found this type of image not only in the first image sequence of the video but also in the last section of the frame sequence. To overcome this scenario, we decided to consider frames from 101-200 to create our feature matrix. So, we calculated histogram values of each frame having the number from 101 to 200. We averaged them to make one observation which defined as input feature vector. Since we created H, S and V color histogram values, we generated three feature vectors from one video. The video and images were processed in Python OpenCV, Matlab and R.

*D. Feature matrix creation*

We found three feature vectors H, S, and V per fingertip video data. Each feature vector contains 256 histogram values since the pixel intensities are from 0 to 255. Note that, the input feature vector is the average of 100 histogram values of three H, S, and V colors. Here, the dimension of each H, S, or V color pixel vector is 1x256. If we consider all three H, S, and V colors, then the length of the input feature vector of one video will be 3x256 = 768. Therefore, we get an input feature vector of size 1x768 combining the three histogram values of H, S and V. We generate a feature matrix with the dimension of 30x768 for 30 videos of the BCW. Similarly, we create an input feature matrix of size 30x768 for the 30 observations of AmaderGram.

Here we give the data extraction and analysis flow for each video step by step.

Step 1: We extract the red, green, and blue (RGB) pixel intensities of 100 image frame. Then we convert these RGB color pixels into HSV color space.

Step 2: We generate 100 histogram values for each H, S, and V color of one video. Make an average of 100 H, S, V color frequencies.

Step 3: We consider averaged H, S and V histogram values of videos as a feature vector. We create one input feature matrix with the dimension of 30x768 from the BCW data. The second input feature matrix is built using the AmaderGram data set that has the size of 30x768.

Step 4: We name the input feature matrix X (later mentioned as predictor matrix X). The clinical hemoglobin levels are stored in Y matrix (called response matrix Y later). This clinical hemoglobin data is considered as gold standard Hb data, and the matrix has the dimension of 30x1 for the BCW and AmaderGram.

Step 5: We standardize each input matrix before data analysis. In this case, each column of X is centered on having mean 0 and scaled to have standard deviation of 1.

Step 6: Partial least squares (PLS) algorithm works better on a large number of feature space [39], [40]. Thus, we apply PLS algorithm on the input feature matrices generated by the BCW and AmaderGram. We use 10-fold cross validation and 10 PLS components in PLS on both data sets.

We present the coefficient of determination ($R^2$) for 10 PLS components in the Result and Analysis section.

VI. RESULT AND ANALYSIS

In this analysis, the variation in multiple pixel values is analyzed by using the Partial Least Squares (PLS) rather than applying the Principal Component Analysis (PCA). PLS offers better accuracy as well as effectiveness in the case of analysis than that of PCA [41] [42]. The input feature matrix built with H, S and V histograms values are drawn on Hb and intensity number (mentioned here as features). The intensity number is from 0 to 767. Here we observe that the frequency of H, S and V pixel intensities change with lower and higher Hb levels. A range of hemoglobin levels is uniformly distributed in this data set as illustrated in Figure 7. The AmaderGram video data produce the feature matrix for 30 subjects which is presented in Figure 8 where the similar level of Hb levels are overlapping.

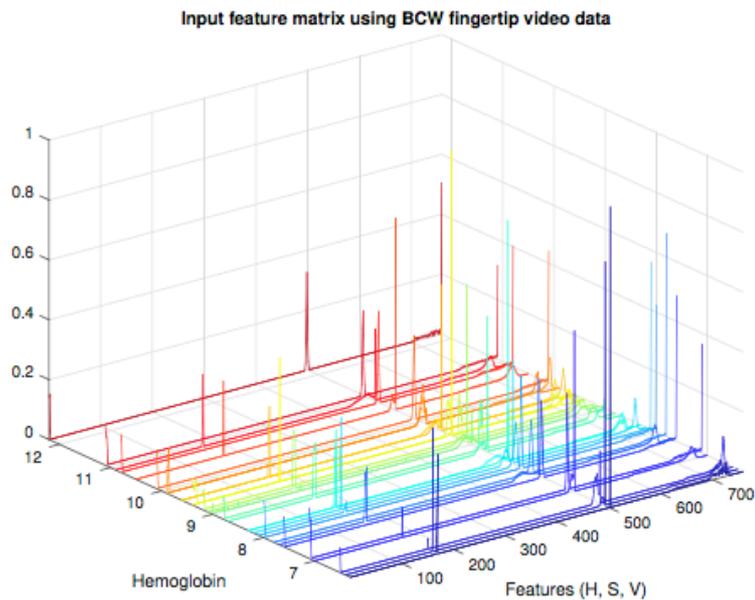

Fig. 7: Input feature matrix for the Blood Center of Wisconsin, USA.

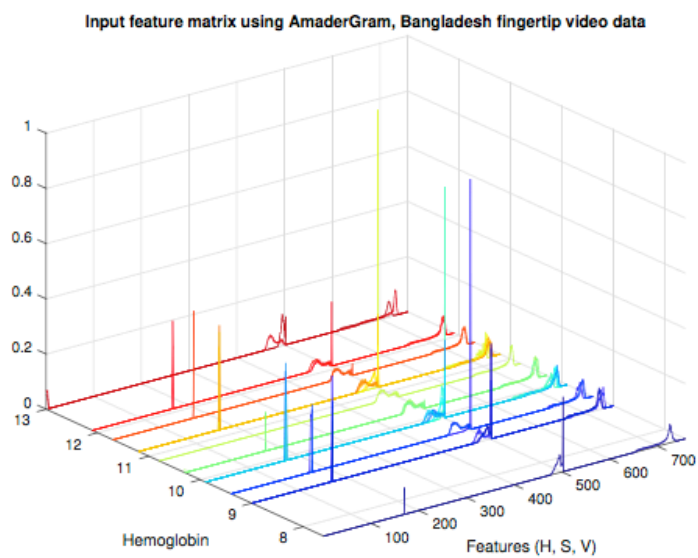

Fig. 8: Input feature matrix for AmaderGram, Bangladesh.

In the case of PLS regression, the $R^2$ test identifies how well the PLS regression model can predict experimental data. The $R^2$ test presents the proportion of variation in the responses comparing to that of the original model. The higher the value of $R^2$, the better the result in the course of Hb level prediction. We do not increase the number of PLS component more than ten since the PLS model accuracy is not going up proportionally. We use 10-fold cross-validation (CV) for each PLS component during data analysis. With 10 PLS components on the BCW data set we get $R^2 = 0.95$. The regression line is presented in Figure 9. Also, the residuals of the BCW model are illustrated in Figure 10.

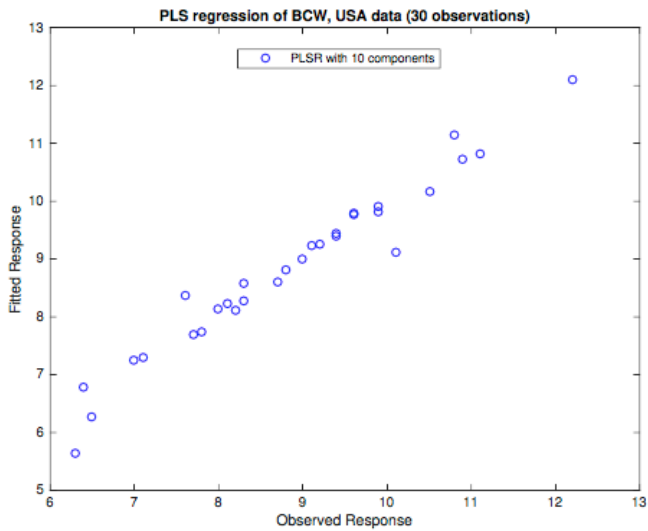

Fig. 9: PLS Regression for 30 subjects using 10 PLS components on the Blood Center of Wisconsin video data.

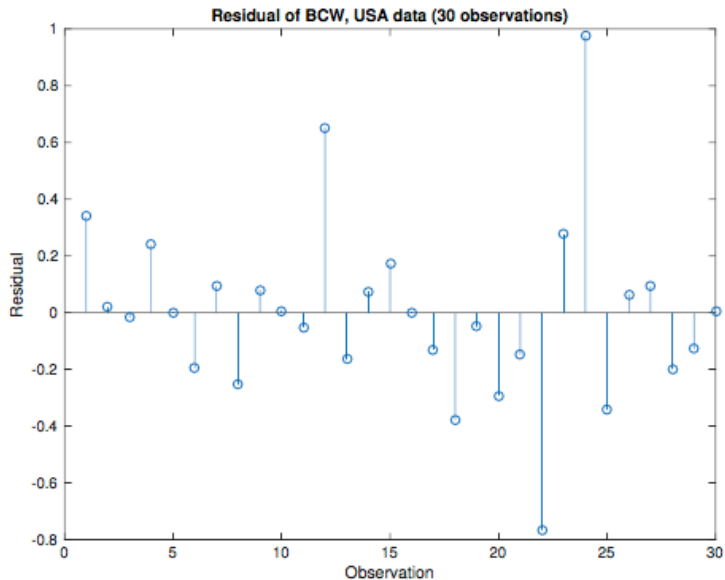

Fig. 10: Residuals of PLS model based on Blood Center of Wisconsin video data.

We analyze the fingertip video data set collected from AmaderGram, Bangladesh. We also choose 10 PLS components and 10-fold cross validation for this data set, and we get $R^2 = 0.95$. The predicted data using the PLS model is shown in Figure 11. The residuals of the PLS model of AmaderGram is presented in Figure 12.

The first objective of the project is to create a ground truth about this experiment. In Section IV-C, we show how the lower-level-hemoglobin-image pixel frequency is changed for the higher-level-hemoglobin-image. We meet this fundamental objective through our investigation that forwards us to do further research on the BCW data first. Later we apply similar approach on AmaderGram, Bangladesh data. In Section V-B, we have presented the novel video collection protocol. We also provide recommendations for the health assistant who will use the smartphone to capture videos. This unique data collection approach satisfies our second objective of this research work. The third objective is to make input feature matrix from the videos only. We present the feature selection process in Section V-D where only H, S and V are used. The frequency of each HSV pixel intensity is captured and is stored as features. This excellent idea works well even if the image resolution is higher in the future. The third objective is achieved here applying our novel techniques.

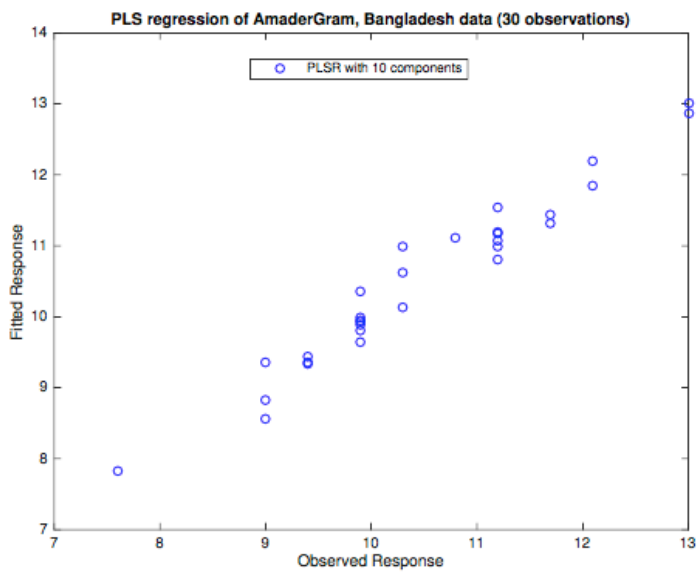

Fig. 11: PLS Regression for 30 subjects with 10 PLS components using AmaderGram video data.

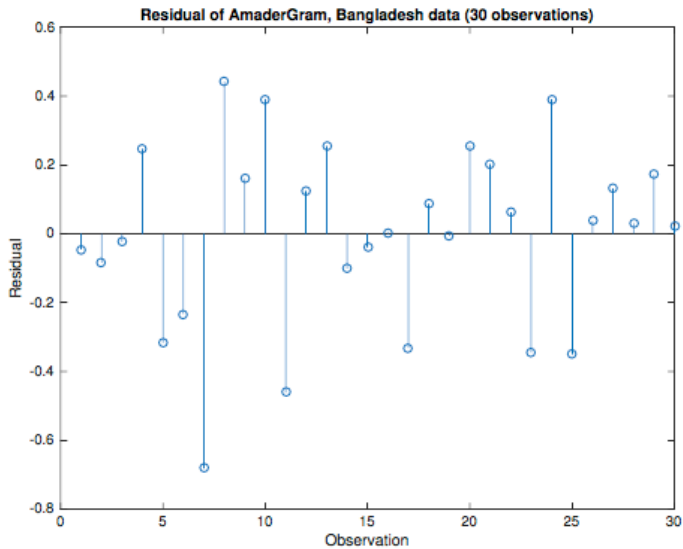

Fig. 12: Residuals of PLS model based on AmaderGram video data.

We take two different data sets to observe how the accuracy and model characteristics are changing by using the novel idea. We found a remarkable change in an output of these two data sets though the skin color issue is an important consideration here. The LED flashlight has high intensity, and it can penetrate easily through different types of skin. So the flashlight removes the skin color issues in this research. We also look for the problems coming from skin thickness, but we observe very few variations in the result. We selected a pretty similar number of male and female subjects from both locations to understand the truth. We saw that HSV color space has the best pixel information for hemoglobin measurement noninvasively.

## VII. CONCLUSIONS

We delineate a noninvasive hemoglobin measurement technique which is based on a smartphone. We use HSV color space of the fingertip video to build the input feature matrix. We have analyzed a severe sickle cell patent's blood transfusion data first. Then we have applied the idea to two different data sets collected from USA and Bangladesh. We use PLS algorithm on both input feature matrices where we found nearly 95% accuracy. We have compared and analyzed the accuracy of these two different data sets. We found that our prediction model provide a reliable result that can be used in the poor and developing countries. In the future, we will increase the number of subjects and features to make a robust hemoglobin measurement model noninvasive way.

## ACKNOWLEDGMENT

The research work is accomplished under a financial support of the Ubicomp Lab, Marquette University and the Blood Center of Wisconsin. The Ubicomp lab at Marquette University specializes in the development of new and innovative ubiquitous and pervasive computing technology, ranging from applications (mobile and desktop) to middleware and even hardware projects. The Blood Center of Wisconsin provides communities with a continuum of care that includes the discovery, diagnosis, treatment and cure of many lifealtering health conditions. The authors are grateful to the Ubicomp Lab at Marquette University, the Blood Center of Wisconsin, and the supporting colleagues and staffs who helped us for the data collection from the medical center.